\documentclass[runningheads]{llncs}
\usepackage{todonotes}
\usepackage{ctable}
\usepackage{xcolor}
\usepackage{mathtools}
\usepackage{multirow}
\usepackage{algorithm}
\usepackage{algorithmic}
\usepackage{adjustbox}
\usepackage{graphicx}
\usepackage{amsfonts}

\newcommand{\edit}[1]{\textcolor{red}{#1}}

\begin{document}
\title{Idle is the New Sleep: Configuration-Aware Alternative to Powering Off FPGA-Based DL Accelerators During Inactivity}
\titlerunning{Idle is the New Sleep: Configuration-Aware Alternative to Powering Off}
\author{Chao Qian\inst{} \and
Christopher Cichiwskyj\inst{} \and
Tianheng Ling\inst{} \and
Gregor Schiele\inst{}}
\authorrunning{C. Qian et al.}
\institute{Intelligent Embedded Systems Lab,\\ 
University of Duisburg-Essen, Germany \\
\email{\{chao.qian, christopher.cichiwskyj, \\ tianheng.ling, gregor.schiele\}@uni-due.de}}


%
\maketitle             

\begin{abstract}

In the rapidly evolving Internet of Things (IoT) domain, we concentrate on enhancing energy efficiency in Deep Learning accelerators on FPGA-based heterogeneous platforms, aligning with the principles of sustainable computing. Instead of focusing on the inference phase, we introduce innovative optimizations to minimize the overhead of the FPGA configuration phase. By fine-tuning configuration parameters correctly,  we achieved a 40.13-fold reduction in configuration energy. Moreover, augmented with power-saving methods, our Idle-Waiting strategy outperformed the traditional On-Off strategy in duty-cycle mode for request periods up to 499.06 ms. Specifically, at a 40 ms request period within a 4147 J energy budget, this strategy extends the system lifetime to approximately 12.39$\times$ that of the On-Off strategy. Empirically validated through hardware measurements and simulations, these optimizations provide valuable insights and practical methods for achieving energy-efficient and sustainable deployments in IoT.

\end{abstract}

\section{Introduction}
\label{sec:introduction}

Embedded Deep Learning (DL) has recently made significant progress in the Internet of Things (IoT) domain~\cite{akkad2023embedded}. Nevertheless, IoT devices, limited by the low-power Microcontroller Units (MCUs), often struggle with performance constraints~\cite{situnayake2023ai}.
Combining Field-Programmable Gate Arrays (FPGAs) with MCUs to create a heterogeneous computing platform has proven to be a promising approach to adapt to these limitations, balancing between computational power and energy efficiency~\cite{krishnamoorthy2021systematic}.
However, optimizing task offloading from MCUs to FPGAs remains crucial to meet the stringent energy budget of IoT devices and advance sustainable IoT ecosystems.

Previous studies typically assumed that there is continuous data or work supply for FPGAs~\cite{gan2021cost,olney2022efficient,qian2022enhancing}, justifying their emphasis on energy efficiency during the inference phase of FPGA-based DL accelerators. However, in common IoT applications like time series analysis, FPGAs often complete inferences much faster than sensor data can be gathered, resulting in a data supply gap.

\begin{figure}[!htb]
\centering
\vspace{-15 pt}
\begin{minipage}{.490\columnwidth}
    \centering
    \includegraphics[width=1\textwidth]{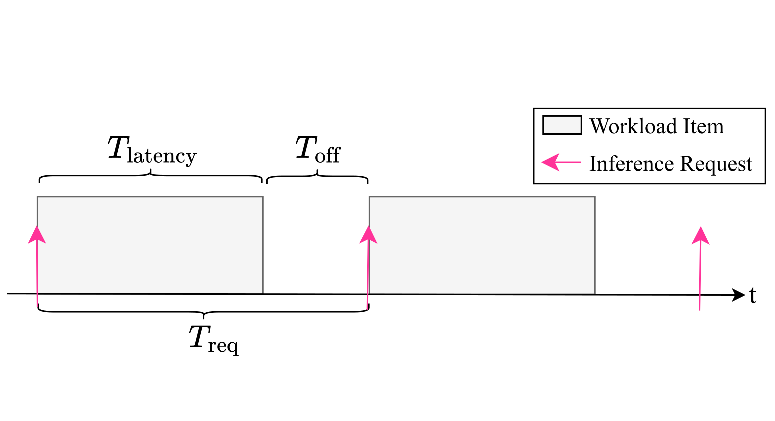}
    \caption{Workloads in Duty-Cycle Mode} 
    \label{fig:workload_item}
\end{minipage}
    \hspace{.01\columnwidth}
\begin{minipage}{.4\columnwidth}
    \centering
    \includegraphics[width=1\textwidth]{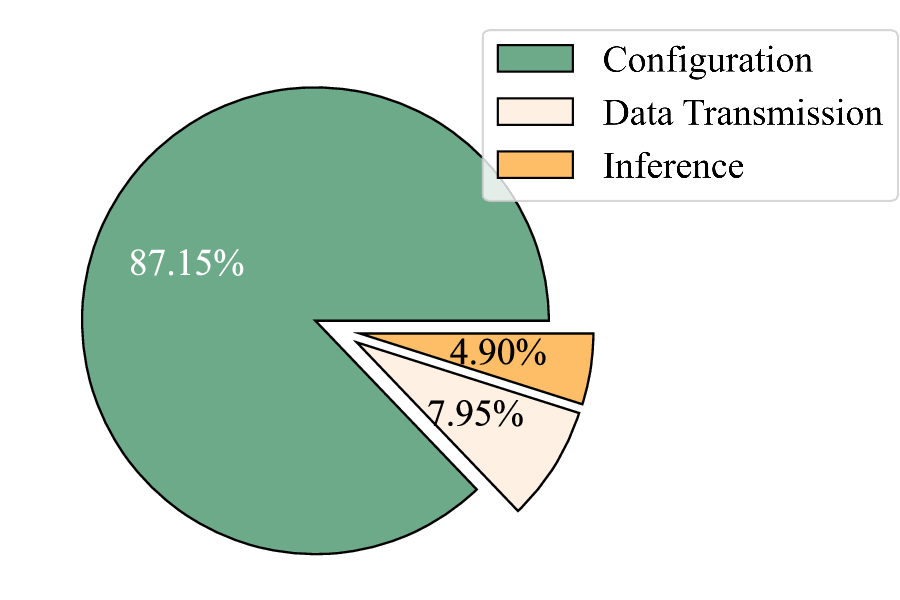}
    \caption{Energy of a Workload Item}
    \label{fig:energy_pie_chart}
\end{minipage}
\end{figure}

In this context, the FPGA can operate in duty-cycle mode~\cite{cheour2019recent}, as shown in Figure~\ref{fig:workload_item}. This mode involves the MCU gathering sufficient data before initiating an inference request (indicated by pink arrows in Figure~\ref{fig:workload_item}) to offload tasks to the FPGA. We term the interval between these requests as the request period (\(T_\text{req}\)), which remains constant in our study that focuses on periodic inference requests. Further, each sequence of operations performed by the FPGA in response to an inference request is defined as a workload item (depicted as a gray box in Figure~\ref{fig:workload_item}), with its processing time labeled as \(T_\text{latency}\). In this study, we explore cases where \(T_\text{latency}<\)\(T_\text{req}\), allowing the FPGA to be powered off after finishing a workload item, thereby conserving energy. The duration of this power-off state is denoted as \(T_\text{off}\).

Each workload item for the DL accelerator involves multiple phases: configuration, data transmission (including loading and offloading), and inference. Our prior research~\cite{cichiwskyj2020time} reveals that the configuration phase accounts for a substantial 87.15\% of the total energy consumption per workload item, as shown in Figure \ref{fig:energy_pie_chart}. This underscores the importance of optimizing this phase to enhance overall energy efficiency. In response, we introduce an incremental approach to reduce the overhead of FPGA configuration.
This approach has been validated through measurements on an FPGA hardware platform and is further supported by simulations for scenarios that could not be achieved by the current design of the hardware platform. The primary contributions of our research are as follows: 

\begin{itemize}
    \item Utilizing a representative embedded \emph{Spartan-7 XC7S15} FPGA, we deliver valuable insights for setting FPGA configuration parameters correctly, reducing the energy consumption by 40.13-fold per FPGA configuration, effectively lowering it to a mere 11.85 mJ. 

    \item We introduce the Idle-Waiting strategy as a more efficient alternative for shorter request periods. In the case of a Long Short-Term Memory (LSTM) accelerator, we demonstrate that for request periods below 89.21 ms within an energy budget of 4147 J, this strategy consistently surpasses the traditional On-Off strategy. At a 40 ms request period, this strategy yields 2.23$\times$ more workload items and a comparable increase in system lifetime.
    
    \item By further integrating two power-saving methods, we reduce FPGA idle power by 81.98\%. This results in a 5.57-fold increase in the number of executable workload items, and the estimated system lifetime can be improved to an average of 47.80 hours within the same energy budget. This optimization also expands the advantageous request period to 499.06 ms.
    
\end{itemize}

The rest of the content is structured as follows:
Section~\ref{sec:system_model} details the system model adopted in our study. Section~\ref{sec:problem_statement} defines the problem we address. Section~\ref{sec:proposed_solution} describes our proposed solutions. Section~\ref{sec:results_evaluation} discusses the experiments and results. Section~\ref{sec:related_work} reviews relevant literature. Finally, Section~\ref{sec:conclusion_future} concludes our research and outlines directions for future work.

\section{System Model}
\label{sec:system_model}

\begin{figure}[!htb]
    \centering
    \vspace{-10 pt}
    \includegraphics[width=0.8\columnwidth]{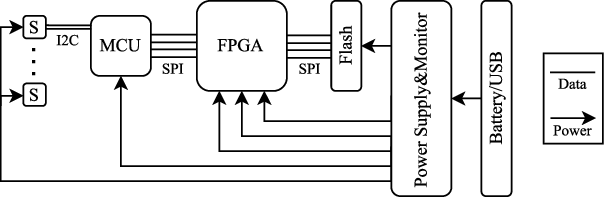}
    \caption{Architecture for the Hardware designed for DL Accelerator}
    \label{fig:system_model}
\end{figure}

This section delves into the system model used in our study, laying the groundwork for understanding the challenges and constraints we address. Figure \ref{fig:system_model} depicts the architecture of the heterogeneous platform we adopted. This platform consists of a low-power \emph{RP2040} MCU for coordination tasks, coupled with an embedded \emph{Spartan-7 XC7S15} FPGA. The MCU is usually in sleep mode, consuming 180 $\mu$A of current. It is woken up by either external hardware interrupts or through timers to perform periodic tasks. Meanwhile, the FPGA serves as a hardware accelerator and is activated only when needed, primarily for processing and accelerating DL tasks.

The communication interface connecting the MCU and the FPGA is a Serial Peripheral Interface (SPI). The FPGA is connected to flash with a dedicated SPI interface, supports clock frequencies from 3 to 66 MHz, and can be programmed to operate in single, dual, or quad buswidths. The FPGA can fetch bitstreams through this interface, facilitating seamless configuration when powering up or switching between different accelerators. 

The power supply of the system is designed in a dual-mode arrangement, consisting of both a USB connection and a 320 mAh rechargeable LiPo Battery. The battery provides an energy capacity of approximately 4147 J, serving as the system's designated energy budget, denoted as \(E_{\text{Budget}}\). To better profile the energy consumers in this system, it includes an energy monitoring subsystem equipped with two PAC1934 sensors, sampling at 1024 times per second for power rail. Figure \ref{fig:system_model} also illustrates the seven monitored power rails that are critical to the system's operation, with lines and arrows marking their interconnections.

\section{Problem Statement}
\label{sec:problem_statement}

As highlighted in Section \ref{sec:introduction}, with periodic inference requests, the FPGA in our system can be powered off for a duration of $T_\text{off}$ to conserve energy. However, for SRAM-based FPGAs like ours, powering off results in the loss of configuration data stored on the chip. This necessitates reconfiguring the FPGA from external flash for each powering on, adding significant overhead to every inference request.

Figure \ref{fig:energy_pie_chart} demonstrates that enhancements in data transmission and inference phases have a limited effect on the overall energy consumption per workload item. Reducing the energy consumption of these phases to zero would only lead to a 12.85\% decrease in the total energy per workload item. In contrast, eliminating the energy overhead associated with the configuration phase could potentially enable the execution of up to 6 additional inference requests, effectively allowing the processing of up to 6$\times$ more workload items within the same energy budget. 

Consequently, our study aims to achieve two primary goals: firstly, to reduce the energy consumed during the configuration phase in one workload item, and secondly, to decrease the number of necessary configurations while considering the request period. 

\section{Proposed Solution}
\label{sec:proposed_solution}

This section details our proposed solutions in three steps. First, we focus on reducing the energy consumption during the configuration phase of a single workload item. Next, we extend our approach to optimize average energy consumption across multiple workload items. Finally, we introduce an analytical model for estimating the executable workload items of each strategy under given application requirements.

\subsection{Reducing Energy for FPGA Configuration Phase}
\label{sec:opt_energy_for_a_infer_req}

\begin{figure}
\vspace{-5 pt}
    \centering
    \includegraphics[width=0.8\textwidth]{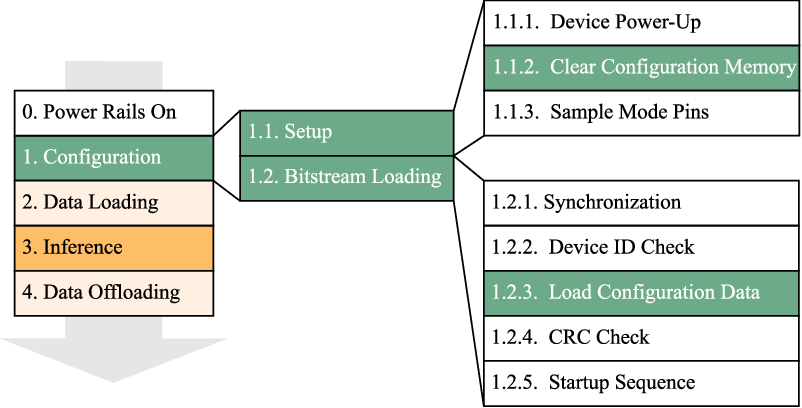}
    \caption{Breakdown of FPGA Configuration Phase}
    \label{fig:configuration_optimization} 
\end{figure}

In the first step, we investigated whether it is possible to reduce or eliminate the energy overhead associated with the FPGA configuration phase by fine-tuning its parameters. 
To achieve this, we delved into the detailed configuration process as outlined in the Xilinx 7-Series FPGA configuration user guide~\cite{amdguide}, mainly focusing on the stages where potential energy savings could be most significant, as shown in Figure \ref{fig:configuration_optimization}. Our empirical analysis highlights that the \emph{Clear Configuration Memory} and \emph{Load Configuration Data} stages are the most energy-intensive. In contrast, other stages contribute minimally to the overall energy consumption due to their short duration.

Our experiments on real hardware show that the \emph{Setup} stage imposes a substantial delay of 27 milliseconds for the \emph{Spartan-7 XC7S15} FPGA after all power rails are ready.
Regrettably, further optimization of this stage proves infeasible due to its inherent dependence on the FPGA model. Due to these constraints, our focus shifts to the \emph{Load Configuration Data} stage. This stage offers three adjustable parameters: SPI buswidth, SPI clock frequency, and bitstream compression Option, as detailed in Table~\ref{tab:bitstream_loading_parameter}. Our approach involves carefully analyzing these parameters to determine the optimal combination that minimizes the energy cost during this stage. Section \ref{subsec:exp1} details the exploration of these parameters, their interplay, and the resulting impact on the energy efficiency of the FPGA configuration phase. 

\begin{table}[!htb]
\centering
\caption{Adjustable Parameters of \emph{Bitstream Loading} Stage on 7-series FPGAs}
\label{tab:bitstream_loading_parameter}
\renewcommand\arraystretch{1.2}
\tabcolsep=0.12cm
\begin{tabular}{|l|l|} 
\hline
\textbf{Parameters} & \textbf{Values} \\ \hline
\textbf{SPI Buswidth} & 1, 2, 4 \\ \hline
\textbf{SPI Clock Frequency} & 3, 6, 9, 12, 16, 22, 26, 33, 40, 50, 66 \\\hline
\textbf{Bitstream Compression Option} & False, True \\ 
\hline
\end{tabular}
\vspace{-10 pt}
\end{table}

While adjusting these parameters yielded a significant 40.13-fold decrease in energy consumption, the configuration phase still imposes a notable energy cost of 11.85 mJ. This observation guides us toward our second solution, which aims to minimize the number of FPGA configurations within the energy budget.

\subsection{Minimize Number of Configurations by Idle-Waiting}
\label{sec:optimise_multiple_requests}

Taking the \emph{Spartan-7 XC7S15} FPGA model as an example, energy consumption in the \emph{Setup} stage is unavoidable. So even if the energy cost of the \emph{Bitstream Loading} stage is optimized to zero, the energy consumption of the configuration phase can only be reduced from 11.85 mJ to 7 mJ. Thus, it becomes imperative to adopt higher-level strategies that account for multiple inference requests to optimize the average energy consumption per inference request. We focus on two strategies that can impact the overhead imposed by the configuration phase. The underlying assumption is that the same accelerator is constantly (re)used for all inference requests. An analysis of supporting different accelerators is outside the scope of this work.

\subsubsection{On-Off Strategy} 
\begin{figure}[!htb]
    \vspace{-15 pt}    
    \centering
    \includegraphics[width=.75\columnwidth]{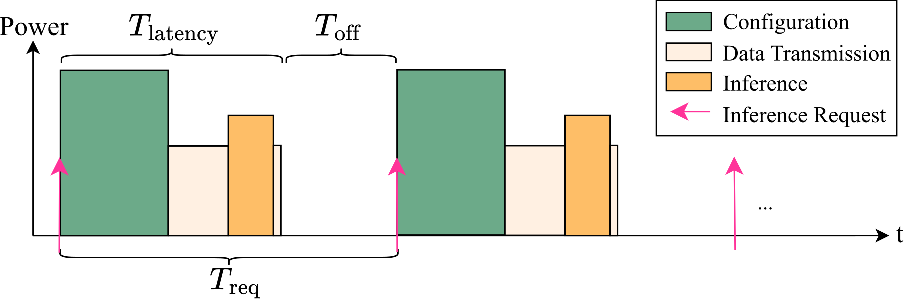}
    \caption{Illustration of On-Off strategy} 
    \label{fig:on_off_strategy}
    \vspace{-10 pt}
\end{figure}

As Section~\ref{sec:introduction} outlines, the FPGA powers off after completing a workload item and reactivates for incoming inference requests. We define this process for periodic inference requests as the On-Off strategy. In this strategy, the FPGA only consumes energy during workload execution, encompassing configuration, data transmission, and inference phases (as depicted in Figure \ref{fig:on_off_strategy}). Notably, the FPGA does not use energy while powered off, and the transition to off state happens instantaneously without energy cost.
Thus, optimizing the energy efficiency involves maximizing the off-time (\(T_\text{off}\)) while maintaining the execution of workload items within the application's latency requirements.
The strategy proves effective when the execution time (\(T_\text{latency}\)) of a workload item is shorter than the request period (\(T_\text{req}\)), allowing sufficient downtime for energy conservation.

\begin{figure}[!htb]
    \vspace{-15 pt}
    \centering
    \includegraphics[width=.75\columnwidth]{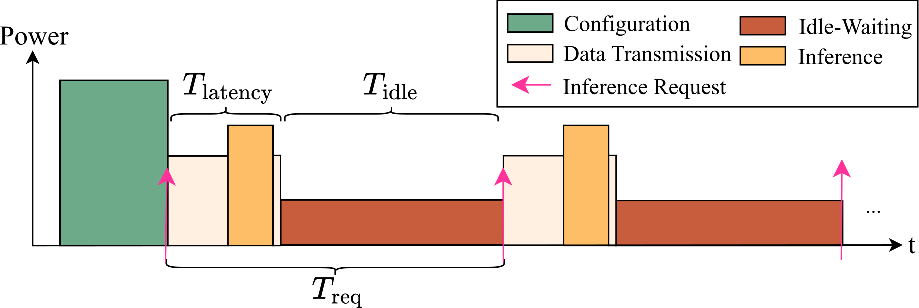}
    \caption{Illustration of Idle-Waiting Strategy}
    \label{fig:idle_waiting_strategy}
    \vspace{-25 pt}
\end{figure}

\subsubsection{Idle-Waiting Strategy}
\label{sec:idle_waiting_strategy}
Extending on the On-Off strategy, we develop an alternative strategy by integrating an idle-waiting phase to replace the conventional powered-off period.
This modification aims to bypass the energy-intensive configuration phase.
Figure~\ref{fig:idle_waiting_strategy} illustrates the core concept of incorporating the idle-waiting phase amidst inference requests. In this strategy, the FPGA undergoes a one-time configuration at the outset, referred to as the initial overhead. Consequently, \(T_\text{latency}\) excludes the time typically consumed by the FPGA configuration phase. This strategy is advantageous in scenarios with frequent inference requests, particularly when the energy overhead of the idle-waiting phase, determined by the time (\(T_\text{idle}\)) of idle-waiting phase and the FPGA's idle power (\(P_\text{idle}\)), is less than the energy consumed in each configuration phase.

To maximize the effectiveness of the Idle-Waiting strategy, we propose two methods to reduce the FPGA's idle power consumption. The first method involves deactivating non-essential components, such as IOs and clock references, when the FPGA is idle. The second method is to reduce the voltage of the FPGA during the idle-waiting phase to a sufficient level to maintain the configuration but not to enable the operational functions, i.e., data transmission and execution of the actual inference. By integrating these two approaches, we anticipate achieving even greater enhancements in reducing power consumption. However, such hardware optimizations necessitate specialized components that may not be readily available in the commercial market. Hence, we employed a simulator to validate and assess the feasibility of these optimizations. The detailed description of this simulator is presented in Section \ref{subsec:experiment_setup}.

\subsection{Analytical Model}
\label{subsec:analytical_model}

To evaluate energy consumption across different strategies within a designated energy budget, we develop an analytical model. This model is instrumental in identifying the maximum number of executable workload items and estimating the corresponding system lifetime.

For the On-Off strategy, \(E_{\text{Sum}}^{\text{OnOff}}(n)\) as outlined in Equation~\ref{eq:total_energy_on_off} represents the cumulative energy cost for \(n\) workload items.
Each \(E_{\text{Item}}^{\text{OnOff}}\) includes the energy consumed during the configuration, data transmission and inference phases. In the Idle-Waiting strategy, as detailed in Equation~\ref{eq:total_energy_idle}, the total energy cost, \(E_{\text{Sum}}^{\text{IdleWait}}(n)\), is comprised of three key components: 
1) \(E_{\text{Init}}\) represents the one-time initial overhead incurred by the FPGA at the start of the system. 2) \(\sum_{i=1}^{n} E_{\text{Item}}^{\text{IdleWait}}\) quantifies the energy required for \(n\) workload items, where all configuration-related overheads are zero. 3) \(\sum_{i=1}^{n-1}E_{\text{Idle}}\) accounts for the energy consumed during idle periods between workload items, where \(E_{\text{Idle}}\) is determined by the idle time (\(T_\text{idle}\)) and the FPGA's idle power consumption (\(P_\text{idle}\)). 

\begin{equation}
E_{\text{Sum}}^{\text{OnOff}}(n) = \sum_{i=1}^{n} E_{\text{Item}}^{\text{OnOff}}
\label{eq:total_energy_on_off}
\end{equation}
\begin{equation}
E_{\text{Sum}}^{\text{IdleWait}}(n) = E_{\text{Init}} + \sum_{i=1}^{n} E_{\text{Item}}^{\text{IdleWait}} + \sum_{i=1}^{n-1}E_{\text{Idle}}
\label{eq:total_energy_idle}
\end{equation}

To ascertain the maximum number (\(n_\text{max}\)) of workload items executable within the energy budget (\(E_{\text{Budget}}\)), we set a criterion ensuring that \(E_{\text{Sum}}(n)\) for different strategies optimally aligns with but does not exceed \(E_{\text{Budget}}\), as formulated in Equation \ref{eq:inference_amount_estimate}. The system lifetime (\(T_\text{lifetime}\)) is then calculated by multiplying the derived (\(n_\text{max}\)) by the request period (\(T_{\text{req}}\)), as per Equation \ref{eq:life_time_estimate}.

\begin{equation}
    n_\text{max} = \max \{ n \in \mathbb{N} \mid E_{\text{Sum}}(n) \leq E_{\text{Budget}} \} 
\label{eq:inference_amount_estimate}
\end{equation}
\begin{equation}
T_\text{lifetime} = n_\text{max} \times T_{\text{req}}
\label{eq:life_time_estimate}
\end{equation}

This analytical model provides a theoretical basis for our research,  enabling a rapid analysis of how various strategies impact energy consumption. We plan to rigorously validate its effectiveness in upcoming experiments, ensuring its practical applicability in real-world scenarios.

\section{Experiments and Results}
\label{sec:results_evaluation}
To validate the effectiveness of our proposed solutions, we conducted three interconnected experiments. The first experiment focused on reducing energy consumption during the FPGA configuration phase. The second experiment assessed the Idle-Waiting strategy's ability to reduce frequent configurations. The third experiment explored the power-saving methods in the idle-waiting phase, further enhancing the strategy's effectiveness.

\subsection{Experiments Setup}
\label{subsec:experiment_setup}

We utilized the hardware specified in Section \ref{sec:system_model} for our experiments. Additionally, to accelerate experimentation and assist in scenarios where direct hardware testing is impractical, we developed a Python-based simulator, inspired by \cite{cichiwskyj2020time}. This tool aligns with the analytical model described in Section \ref{subsec:analytical_model}, and outputs the maximum number of executable workload items along with estimations of the system lifetime.

This simulator enables the specification of overall workload and individual workload items using YAML files, simplifying the execution of extensive experiments involving large datasets or complex measurements. A key feature of this simulator is its ability to incorporate both datasheet specifications and real hardware measurement, thus enhancing the precision of energy consumption estimations and offering a more realistic representation of actual scenarios.

The simulator requires two descriptions to operate: 1) the \emph{workload} and 2) the \emph{workload item}. The \emph{workload} description contains the energy budget \(E_{\text{Budget}}\) in joules and the constant request period, as mentioned in Section \ref{sec:introduction}. The \emph{workload item} description details each phase's average power consumption in milliwatts and duration in milliseconds. With these inputs, the simulator can effectively model the various strategies discussed in Section \ref{sec:optimise_multiple_requests}, allowing us to examine them under diverse conditions.

\subsection{Experiment 1: Optimization on Energy for FPGA Configuration}
\label{subsec:exp1}

In the first experiment, we conducted a hardware-based investigation of various FPGA configuration parameters involving 11 SPI clock frequencies, 3 SPI buswidths, and the bitstream compression option, listed in Table~\ref{tab:bitstream_loading_parameter}. We aimed to assess their impact on energy consumption during the FPGA configuration phase. Utilizing an LSTM accelerator with a hidden size of 20, as detailed in our prior study~\cite{qian2023energy}, we generated corresponding bitstreams for the \emph{Spartan-7 XC7S15} FPGA. The evaluation metrics include configuration time (milliseconds), power usage (milliwatts), and energy consumption (millijoules) during the configuration phase.

\begin{figure}[!hbt]
    \vspace{-15 pt}
    \centering
    \includegraphics[width=1\columnwidth]{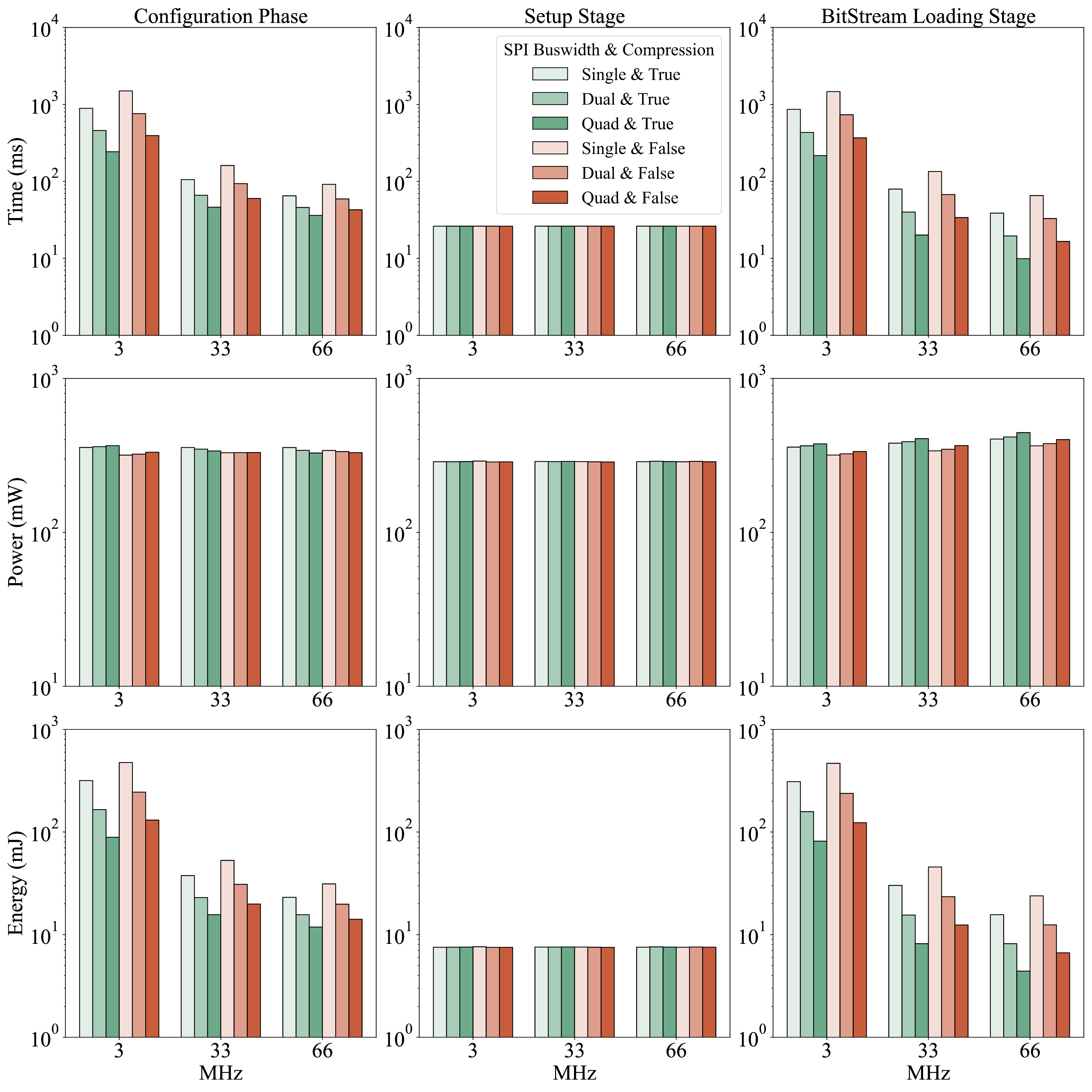}
    \caption{Performance Comparison on the \emph{Spartan-7 XC7S15} FPGA}
    \label{fig:S15_comparison}
    \vspace{-15 pt}
\end{figure}

Due to space constraints, Figure \ref{fig:S15_comparison} selectively presents our results at SPI clock frequencies of 3, 33, and 66 MHz. This selection of data points illustrates the effects across low, medium, and high-frequency settings. The first column of graphs displays the configuration phase outcomes, while the second and third columns break down these results into the \emph{Setup} and \emph{Bitstream Loading} stages, respectively. To facilitate a clearer understanding, the evaluation metrics are displayed in separate rows. The y-axis of each metric is scaled logarithmically.

Our findings indicate a marked decrease in configuration time with increased buswidth and clock frequency.
Employing bitstream compression further enhanced efficiency. In the case of Quad SPI at 66 MHz with compression enabled, it reduced configuration time to 36.15 ms, a 41.4-fold improvement over the least efficient setting of Single SPI at 3 MHz without compression. This decrease is mainly attributed to the impact of settings on the \emph{Bitstream Loading} stage.

The average power consumption during the configuration phase varied, reflecting an aggregate of the \emph{Setup} and \emph{Bitstream Loading} stages. The \emph{Setup} stage maintained a consistent power consumption of around 288 mW. In contrast, the \emph{Bitstream Loading} stage showed increased power usage, especially with larger SPI buswidth and higher frequencies. Notably, bitstream compression led to higher power in this stage, likely due to more switching activities on the SPI data line.

Using Quad SPI at 66 MHz with compression, the energy consumption during the configuration phase was 11.85 mJ, compared to 475.56 mJ for Single SPI at 3 MHz without compression, illustrating a significant 40.13-fold reduction in energy achieved with optimal settings. The trend in energy savings mirrored the timing pattern, with the \emph{Bitstream Loading} stage contributing significantly. Higher SPI frequencies and wider buswidths led to lower energy costs, attributable to the static power characteristics of Spartan-7 FPGAs. Accelerating the bitstream loading process shortened the duration of static power draw, thereby decreasing overall energy consumption.

Similar experiments on the larger \emph{Spartan-7 XC7S25} FPGA yielded comparable results. With the optimal settings, the configuration time for the same accelerator was 38.09 ms, and energy consumption was 13.75 mJ. These results suggest that the highest clock frequency and widest SPI buswidth optimize configuration energy, as long as the hardware supports these settings. However, such settings increase power consumption during the \emph{Bitstream Loading} stage, requiring a higher power budget for the hardware.

\subsection{Experiment 2: Idle-Waiting vs On-Off Strategies}
\label{subsec:exp2}

In this experiment, we set out to identify the request period range where the Idle-Waiting strategy is more efficient than the On-Off strategy. Additionally, we aimed to validate the effectiveness of our analytical model using these experimental results. Utilizing the LSTM accelerator described earlier, we measured timing and power consumption to characterize a workload item, as listed in Table \ref{tab:exp2}. We applied the optimal settings identified in Experiment 1 for the configuration phase. Note that the idle power consumption of 134.3 mW listed in Table \ref{tab:exp2} is specific to the Idle-Waiting strategy. Profiling other accelerators is also feasible, simply requiring an adjustment of the characteristics listed in Table \ref{tab:exp2} to align with the specific accelerator being used.
\begin{table}[!htb]
\renewcommand\arraystretch{1.1} 
\tabcolsep=0.15cm 
\caption{Power and Time on Hardware for Simulation}
\label{tab:exp2}
\centering
\begin{tabular}{|c|c|c|c|}
\hline
\multicolumn{2}{|c|}{\multirow{2}{*}{\textbf{Settings}}} & \multicolumn{2}{c|}{\textbf{LSTM Accelerator\cite{qian2023energy}}} \\ \cline{3-4}
\multicolumn{2}{|c|}{} & \textbf{Power (mW)} & \textbf{Time (ms)} \\ \hline
\multirow{5}{*}{\textbf{Phases}} & \textbf{Configuration}   & 327.9     & 36.145 \\ \cline{2-4}
                                 & \textbf{Data Loading}    & 138.7     & 0.0100  \\ \cline{2-4}
                                 & \textbf{Inference}       & 171.4$^*$  & 0.0281  \\ \cline{2-4}
                                 & \textbf{Data Offloading} & 144.1      & 0.0020  \\ \cline{2-4}
                                 & \textbf{Idle-Waiting}    & 134.3      & varying$^\dagger$ \\
\hline
\end{tabular}\\
\vspace{5 pt}
\small $\dagger$ The idle time varies in response to changes in the request period.\\
\small $*$ This power includes the 114 mW for clock reference and flash chip.
\vspace{-10 pt}
\end{table}
\begin{figure}[!htb]
\centering
\begin{minipage}{.485\columnwidth}
    \centering
    \includegraphics[width=1\textwidth]{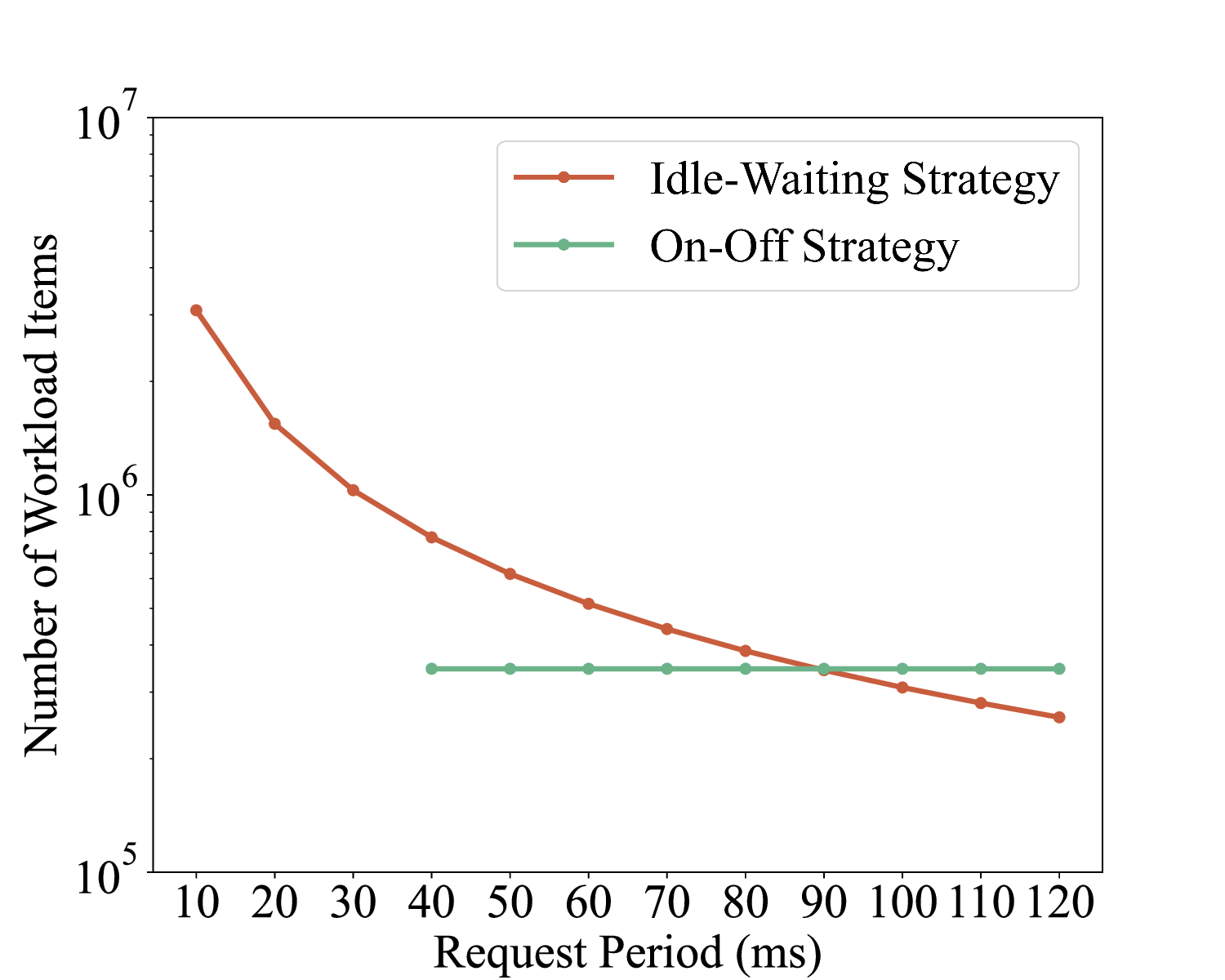}
    \caption{Workload Items: Idle-Waiting vs On-Off Strategies} 
    \label{fig:exp2_simulator_total_number}
\end{minipage}
    \hspace{.005\columnwidth}
\begin{minipage}{.485\columnwidth}
    \centering
    \includegraphics[width=1\textwidth]{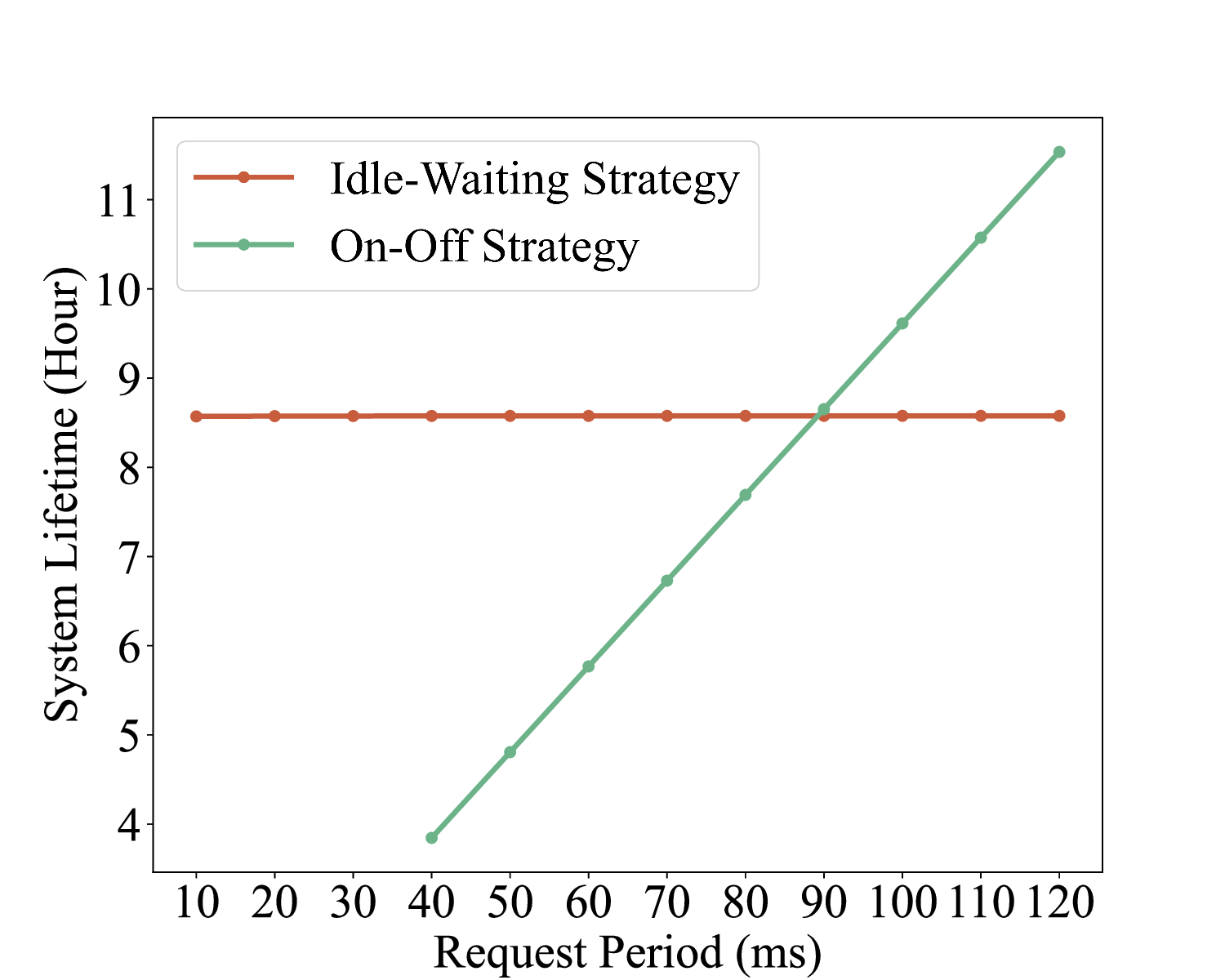}
    \caption{System Lifetime: Idle-Waiting vs On-Off Strategies}
    \label{fig:exp2_simulator_lifetime}
\end{minipage}
 \vspace{-10 pt}
\end{figure}

Utilizing the simulator, we first estimated the number of executable workload items within an energy budget of 4147 J for request periods ranging from 10 to 120 ms, in increments of 0.01 ms. This range was chosen to align with our analytical model's prediction of an efficiency cross point around 89.21 ms, enabling a comprehensive performance analysis.

Figure~\ref{fig:exp2_simulator_total_number} displays the number of executable workload items on a logarithmic scale. For brevity, we show values at 10 ms intervals. Under the Idle-Waiting strategy, the number of executable workload items ranges from a minimum of approximately 257,305 to a maximum of about 3,085,319. Conversely, the On-Off strategy consistently supports 346,073 executable items. At a 40 ms request period, the Idle-Waiting strategy yields 2.23$\times$ more workload items. Note that the On-Off strategy is not represented for request periods below 36.15 ms due to its configuration time cost. When the request period is shorter than 36.15 ms, the FPGA can not be prepared to process an incoming workload. The cross point at 89.21 ms shown in Figure~\ref{fig:exp2_simulator_total_number}, as identified by our analytical model, highlights the effectiveness of the Idle-Waiting strategy for request periods shorter than this threshold.

Figure \ref{fig:exp2_simulator_lifetime} provides an estimation of the system lifetime under each strategy. For the Idle-Waiting strategy, the lifetime averages about 8.58 hours, with a marginal increase across different request periods. This subtle variation, while not prominently visible in the figure, aligns with expectations: the average power consumption in the Idle-Waiting strategy tends to approach idle power levels, since the other phases consume very little energy for their duration. The cross point in Figure~\ref{fig:exp2_simulator_lifetime} mirrors the findings in Figure~\ref{fig:exp2_simulator_total_number}. In contrast, the On-Off strategy exhibits a linear increase in system lifetime as request periods extend.

To assess the precision of our simulator, we conducted direct hardware measurements at a 40 ms request period for both strategies, considering the FPGA and its peripherals' energy consumption. These measurements exhibited only minor variations, with a 2.8\% difference in executable workload items and a 2.7\% discrepancy in system lifetime, thereby affirming the simulator's usability in offering preliminary insights into the accelerator's behavior.

Our results affirm the potential of the Idle-Waiting strategy for request periods below 89.21 ms. The main limitation arises from the power consumption during the idle-waiting phase. We believe that reducing idle power could allow for the execution of more workload items, thereby expanding the range of request periods where the Idle-Waiting strategy is effective.

\subsection{Experiment 3: Optimization on the Idle-Waiting Strategy}
\label{subsec:exp3}

In the final experiment, we aimed to improve the Idle-Waiting strategy by reducing idle power consumption, as proposed in Section \ref{sec:optimise_multiple_requests}. We evaluated these enhancements against the initial Idle-Waiting strategy (referred to as Baseline) from Section~\ref{subsec:exp2}. We utilized the hardware setting described in Section \ref{sec:system_model} to assess the feasibility of these methods, and the results are detailed in Table \ref{tab:exp3}.

\begin{table}[htb]
\renewcommand\arraystretch{1.1} 
\tabcolsep=0.15cm 
\vspace{-10 pt}
\caption{Idle Power on Hardware for Simulation}
\label{tab:exp3} 
\centering 
\begin{tabular}{|c|c|c|c|}
\hline
\multirow{2}{*}{\textbf{Metric}} & \multicolumn{3}{c|}{\textbf{Optimization Methods}} \\ \cline{2-4}
                                 & \textbf{Baseline} & \textbf{Method 1} & \textbf{Method 1+2} \\ \hline
\textbf{Idle Power (mW)}  & 134.3            & 34.2            & 24.0             \\ \hline
\textbf{Saved Power (\%)}        & -                 & 74.38           & 81.98              \\ \hline
\end{tabular}
\vspace{-10 pt}
\end{table}

By deactivating the clock reference and FPGA IOs (Method 1), the idle power can be reduced to 34.2 mW, achieving a 74.38\% power saving compared to the baseline. Further reductions were achieved by lowering the FPGA's internal and auxiliary supply voltages (Method 2) from 1.0 V and 1.8 V to 0.75 V and 1.5 V, respectively. Combining Methods 1 and 2 lowered the idle power to 24 mW, resulting in an 81.98\% reduction compared to the baseline. It is important to note that our hardware setup includes a flash component with a constant power consumption of approximately 15.2 mW, which is factored into all power calculations presented in Table~\ref{tab:exp3}. We verified on our hardware that exiting from these power-saving methods does not affect the FPGA's configuration, ensuring it remains retained and functional. 

\begin{figure}[!htb]
    \vspace{-5 pt}
    \centering
    \includegraphics[width=.89\textwidth]{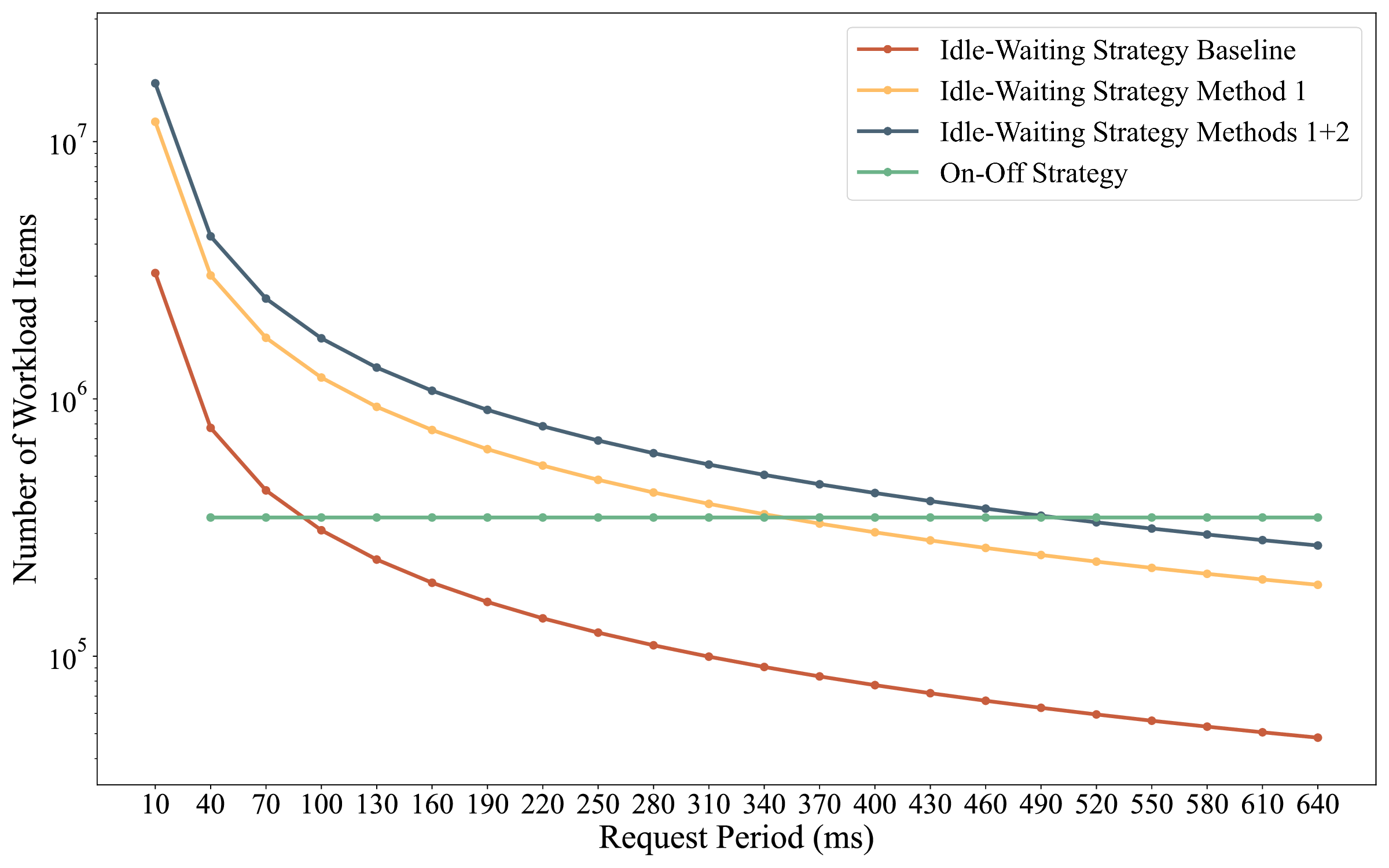}
    \caption{Workload Items: Baseline vs. Optimized Methods Across Request Periods}
    \label{fig:exp3_simulator_total_number}
\end{figure}

\begin{figure}[!htb]
    \centering
    \includegraphics[width=.89\textwidth]{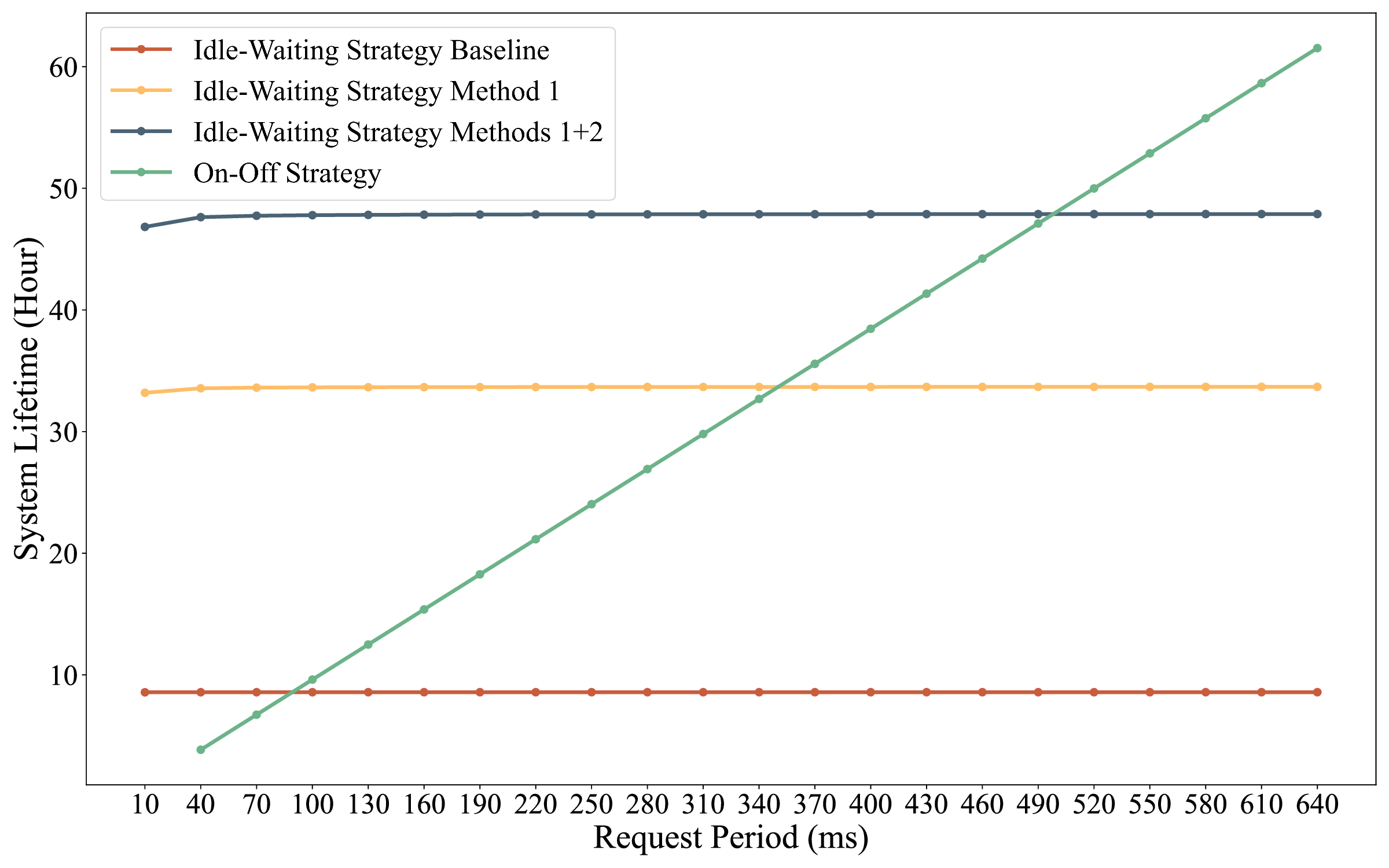}
    \caption{System Lifetime: Baseline vs. Optimized Methods Across Request Periods}
    \label{fig:exp3_simulator_lifetime}
    \vspace{-10 pt}
\end{figure}

Due to the lack of dynamic voltage scaling support in our hardware, we relied on our simulator to estimate potential improvements for test cases that the hardware could not support.
Figure \ref{fig:exp3_simulator_total_number} illustrates that Method 1 alone significantly enhanced the number of executable workload items by 3.92$\times$ relative to the Baseline. Combining both methods yielded even more substantial gains, increasing the number to 5.57$\times$ the Baseline. Consequently, Figure \ref{fig:exp3_simulator_lifetime} shows a proportional increase in system lifetime. Applying method 1 extended the average lifetime to 33.64 hours, a 3.92-fold improvement from Baseline. Furthermore, the combination of methods 1 and 2 further augmented the average lifetime to 47.80 hours. These enhancements expanded the beneficial request period range for the Idle-Waiting strategy from 89.21 ms to 499.06 ms.

The effectiveness of our power-saving methods in increasing executable workload items is evident, yet further idle power reduction is needed. Hardware constraints, mainly the flash's power consumption, limit our method's optimal period to 499.06 ms. Addressing this could extend the advantageous period by up to 5.57$\times$.

\section{Related Work}
\label{sec:related_work}
 
Recent research on DL accelerators on FPGAs has made significant progress, particularly in throughput and energy efficiency~\cite{magyari2022review,muralidhar2022energy,qian2022enhancing}. These studies typically focus on specialized hardware design and optimized inference phase to boost accelerator performance, primarily during continuous processing tasks, with the FPGA constantly active. However, these studies frequently neglect the overhead of FPGA configuration and related phases. For example, Chen et al.~\cite{chen2021eciton} focused on executing a single inference after a power failure, assuming that the FPGA is pre-configured before the power failure. Thus, they only need to optimize the performance of the accelerator in the inference phase. In practical scenarios, DL tasks typically demand a series of inferences, not just a single inference execution.

Some researchers have started to address the impact of FPGA configuration overhead on efficiency. Fritzsch et al.~\cite{fritzsch2022reduction} proposed a method to compress the bitstream by 1.05 to 12.2$\times$ to reduce configuration time, yet they did not explore its energy efficiency implications.
Cichiwskyj et al.~\cite{cichiwskyj2020time} introduced the concept of Temporal Accelerators, demonstrating that even with two reconfigurations, using a smaller FPGA (\emph{Spartan-7 XC7S6}) is more energy-efficient than a larger one (\emph{Spartan-7 XC7S15}) for a single inference execution. However, these studies did not integrate the configuration overhead with request period considerations for energy efficiency.

Our study differs by aiming to enhance the energy efficiency of embedded DL systems through the lens of periodic workload requests. We adopted a dual-phase approach: firstly, we optimized configuration parameters to minimize overhead, and secondly, we employed idle power optimizations to maintain the FPGA powered on, thus avoiding repeated configurations. 


\section{Conclusion and Future Work}
\label{sec:conclusion_future}

In conclusion, this study significantly reduced the configuration overhead of FPGA-based DL accelerators in IoT applications.
By optimizing the FPGA configuration phase and introducing an effective Idle-Waiting strategy, we demonstrated substantial energy savings, thereby increasing the number of executable workload items.
Our Idle-Waiting strategy effectively addresses the challenges of shorter request periods, a limitation of the traditional On-Off strategies.
Combining the Idle-Waiting strategy with idle power-saving methods at a 40 ms request period, it achieves 12.39$\times$ more workload items and system lifetime than the On-Off strategy.

In the future, we plan to extend our power-saving techniques beyond DL use cases to other periodic processes. Additionally, we aim to investigate methods for efficiently handling irregularly occurring inference requests, focusing on optimizing energy efficiency and system performance in more complex scenarios. These efforts will help validate and refine our approaches in diverse operational contexts.

\noindent{\textbf{Acknowledgements.}} The authors acknowledge the financial support provided by the Federal Ministry for Economic Affairs and Climate Action of Germany in the RIWWER project (01MD22007C).

\bibliographystyle{splncs04}
\bibliography{references}

\end{document}